\documentclass[conference]{IEEEtran}
\IEEEoverridecommandlockouts

\makeatletter
\newcommand{\linebreakand}{%
  \end{@IEEEauthorhalign}
  \hfill\mbox{}\par
  \mbox{}\hfill\begin{@IEEEauthorhalign}
}
\makeatother

\usepackage{comment}
\usepackage{float}
\usepackage{cite}
\usepackage{amsmath,amssymb,amsfonts}
\usepackage{algorithmic}
\usepackage{graphicx}
\usepackage{textcomp}
\usepackage{xcolor}
\def\BibTeX{{\rm B\kern-.05em{\sc i\kern-.025em b}\kern-.08em
    T\kern-.1667em\lower.7ex\hbox{E}\kern-.125emX}}
\begin{document}

\title{Carbon-Aware Governance Gates: An Architecture for Sustainable GenAI Development}

\author{
\IEEEauthorblockN{Mateen A. Abbasi}
\IEEEauthorblockA{\textit{University of Jyväskylä}\\
Jyväskylä, Finland\\
\texttt{mateen.a.abbasi@jyu.fi}}
\and
\IEEEauthorblockN{Tommi J. Mikkonen}
\IEEEauthorblockA{\textit{University of Jyväskylä}\\
Jyväskylä, Finland\\
\texttt{tommi.j.mikkonen@jyu.fi}}
\and
\IEEEauthorblockN{Petri J. Ihantola}
\IEEEauthorblockA{\textit{University of Jyväskylä}\\
Jyväskylä, Finland\\
\texttt{petri.j.ihantola@jyu.fi}}
\linebreakand

\IEEEauthorblockN{Muhammad Waseem}
\IEEEauthorblockA{\textit{Tampere University}\\
Tampere, Finland\\
\texttt{muhammad.waseem@tuni.fi}}
\and
\IEEEauthorblockN{Pekka Abrahamsson}
\IEEEauthorblockA{\textit{Tampere University}\\
Tampere, Finland\\
\texttt{pekka.abrahamsson@tuni.fi}}
\and
\IEEEauthorblockN{Niko K. Mäkitalo}
\IEEEauthorblockA{\textit{University of Jyväskylä}\\
Jyväskylä, Finland\\
\texttt{niko.k.makitalo@jyu.fi}}

}

\maketitle

\begin{abstract}

The rapid adoption of Generative AI (GenAI) in the software development life cycle (SDLC) increases computational demand, which can raise the carbon footprint of development activities. At the same time, organizations are increasingly embedding governance mechanisms into GenAI-assisted development to support trust, transparency, and accountability. However, these governance mechanisms introduce additional computational workloads, including repeated inference, regeneration cycles, and expanded validation pipelines, increasing energy use and the carbon footprint of GenAI-assisted development. This paper proposes Carbon-Aware Governance Gates (CAGG), an architectural extension that embeds carbon budgets, energy provenance, and sustainability-aware validation orchestration into human–AI governance layers. CAGG comprises three components: (i) an Energy and Carbon Provenance Ledger, (ii) a Carbon Budget Manager, and (iii) a Green Validation Orchestrator, operationalized through governance policies and reusable design patterns.

\end{abstract}
\begin{IEEEkeywords}
Software architecture, AI governance, Generative AI, Sustainable software engineering, Carbon-aware computing, DevOps governance, Energy-efficient systems.
\end{IEEEkeywords}

\section{Introduction}

The growing computational demands of AI systems have intensified concerns about environmental sustainability \cite{zhuk2023artificial}. As AI is increasingly adopted in software engineering, organizations are paying closer attention to the energy use and carbon footprint associated with AI-assisted development workflows \cite{cruz2025greening}. While substantial effort has gone into improving model efficiency and infrastructure utilization, the environmental impacts of end-to-end AI-enabled software development processes remain comparatively underexplored.


Governance mechanisms are increasingly embedded in GenAI-assisted software development workflows to support trust, accountability, transparency, and regulatory compliance \cite{constantinides2024implications}. In practice, this governance introduces validation and oversight steps such as policy checks, auditing, traceability, and iterative regeneration when outputs fail quality or compliance criteria, which collectively add computational overhead. While these mechanisms are essential for responsible AI use, they can increase the energy demand and carbon footprint of the development environment.


Generative AI is transforming software engineering by automating tasks such as code generation, test creation, and documentation. Empirical studies indicate that AI pair-programming tools can improve developer productivity in certain programming activities \cite{peng2023impact}. However, integrating GenAI into development pipelines introduces governance challenges, including unverifiable outputs, limited traceability, and accountability risks. As a result, governance and validation layers are increasingly embedded to enforce oversight, policy compliance, and human review in line with responsible AI principles \cite{floridi2018ai4people, smuha2019eu}. These governance mechanisms add computational overhead. Large language model inference is energy-intensive, and repeated validation and regeneration cycles further increase energy use and the carbon footprint of AI-assisted development \cite{strubell2019energy, patterson2021carbon}. Strengthening governance without considering sustainability may therefore amplify environmental impact.

Research in sustainable AI has largely focused on model optimization, data center efficiency, and training-phase emissions. In contrast, AI governance research emphasizes risk management, compliance, and oversight. The sustainability implications of governance-driven GenAI-assisted software development remain insufficiently explored. 

To address the above \textbf{gap}, we \textbf{propose} Carbon-Aware Governance Gates (CAGG), a layered architectural approach that integrates sustainability considerations into AI governance for GenAI-enabled software development. The architecture extends human–AI governance layers with carbon-aware validation orchestration, emissions budgeting, and provenance observability, enabling governance processes to operate within defined sustainability boundaries.

\section{Background}

The importance of transparency, accountability, and human involvement in AI governance architectures is emphasized in various AI governance frameworks. Floridi et al. have proposed ethical governance principles that include explicability and responsibility \cite{floridi2018ai4people}. The European Commission’s Ethics Guidelines for Trustworthy AI define requirements such as human agency, auditability, and technical robustness \cite{smuha2019eu}.
These frameworks establish governance objectives but do not address sustainability impacts introduced by governance processes themselves.

The carbon footprint of an AI system has already been extensively studied in literature. Strubell et al. quantified the emissions associated with training deep natural language processing models \cite{strubell2019energy}. Patterson et al. have shown that the carbon footprint of an AI system can vary substantially depending on the hardware efficiency and energy sources\cite{patterson2021carbon}.This paper focuses on the efficiency of the infrastructure and the training processes rather than development governance workflows.

Software design decisions influence lifecycle energy consumption \cite{atadoga2024advancing}. Architectural choices therefore shape system-level environmental impact over time \cite{dutta2023case, kruchten2012technical}. However, sustainability research has largely focused on runtime systems rather than AI-enabled development governance. Carbon-aware workload scheduling has been explored in data-center contexts. Radovanovic et al. demonstrated that shifting compute workloads on carbon intensity in grids can lead to significantly reduced emissions\cite{radovanovic2022carbon}.
This approach has not been systematically applied in AI governance or validation workflows.

\section{Carbon-Aware Governance Gates}

\begin{figure}[H]
\centering
\includegraphics[width=0.99\columnwidth]{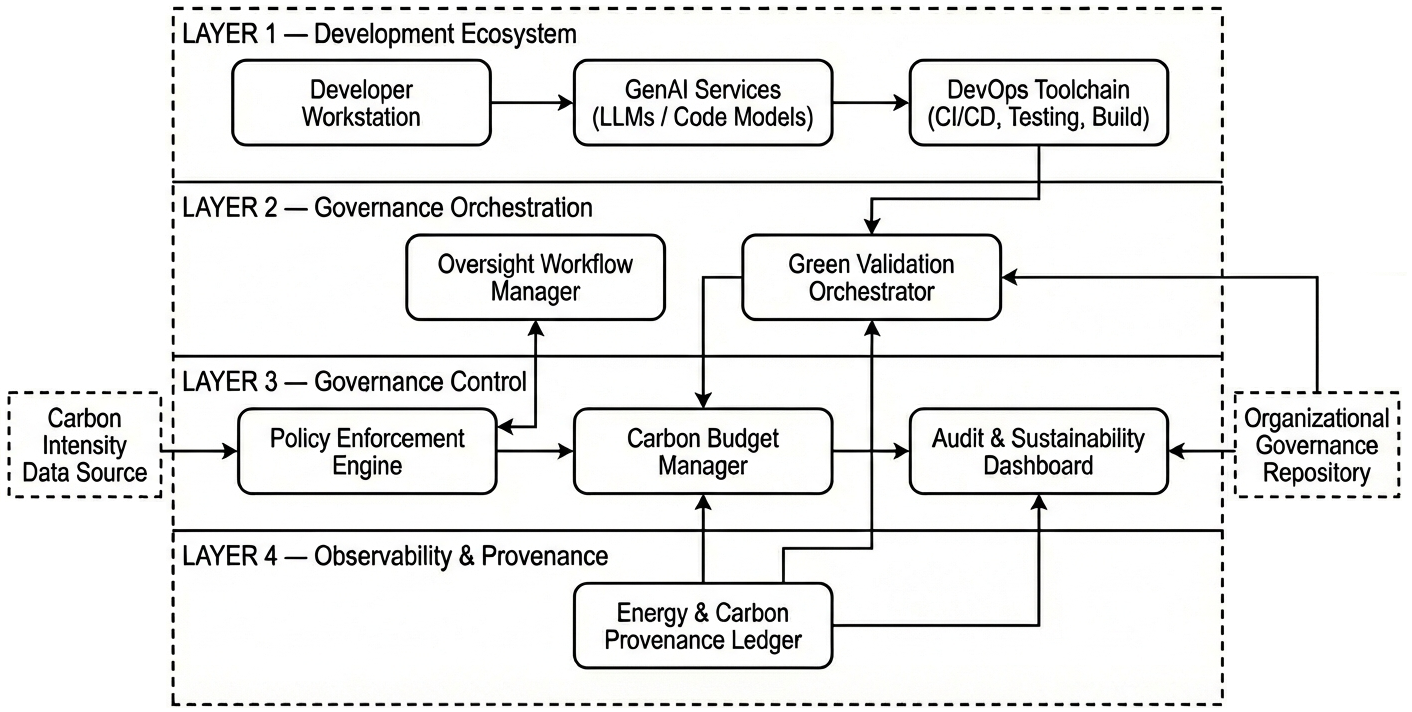}
\caption{Carbon-Aware Governance Layer for GenAI-Enabled Software Development}
\label{fig:cagg_architecture}
\end{figure}

In this section, we introduce Carbon-Aware Governance Gates (CAGG), governance checkpoints integrating sustainability metrics and policies into validation and oversight decisions.
Figure~\ref{fig:cagg_architecture} illustrates the proposed Carbon-Aware Governance (CAG) Layer and its integration with GenAI services and DevOps toolchains.
The figure presents a layered reference architecture for GenAI-enabled software development. Existing governance elements such as the oversight workflow manager and policy enforcement mechanisms are treated as baseline governance infrastructure, whereas CAGG introduces the Green Validation Orchestrator, Carbon Budget Manager, and Energy and Carbon Provenance Ledger as sustainability-specific architectural extensions.


The Development Ecosystem layer captures the boundaries of the development environment in which software development teams interact with GenAI-based services and DevOps toolchains. The oversight workflow manager and green validation orchestrator mediate governance orchestration within GenAI-based software development processes. At the Governance Control layer, policy enforcement mechanisms and carbon budgeting mechanisms are used to manage GenAI-based software development processes in accordance with organizational sustainability objectives. A closed feedback loop connects validation, budgeting, and policy enforcement mechanisms to determine governance decisions. Finally, the Observability and Provenance layer ensures traceability within GenAI-based software development processes through an energy and carbon provenance ledger.

\subsection{Architectural Extensions}

Architectural extensions refer to the sustainability-oriented extensions made to the existing AI governance reference architecture for the purpose of carbon-aware validation, policy enforcement, and observability, while avoiding any changes to the existing governance control structures. While existing AI governance architectures offer the necessary mechanisms for policy enforcement, oversight, and validation orchestration, the sustainability aspects of the governance activities are not explicitly considered. The validation, regeneration, and testing activities performed due to the governance activities impose considerable computational overhead, thus necessitating the need for sustainability-aware control mechanisms.

To address this gap, we propose three architectural extensions that operationalize sustainability governance without restructuring the base governance model. These extensions enable emissions observability, budget regulation, and carbon-aware validation enforcement. The ledger records estimated emissions for each inference and validation stage together with the model identity, execution context, and regeneration count; these records are aggregated into pull-request, pipeline and release-level budgets.

\begin{itemize}
    \item \textbf{Energy \& Carbon Provenance Ledger:} Captures sustainability metadata for inference events, regeneration cycles, and validation workloads using established carbon accounting approaches \cite{patterson2021carbon}.
    \item \textbf{Carbon Budget Manager:} Defines and enforces carbon budgets at pull request, pipeline, and release levels.
    \item \textbf{Green Validation Orchestrator:} This part of the system aims to optimize the validation depth, the models chosen, and the execution schedules. This aligns with the evidence of the influence of software processes on energy consumption \cite{atadoga2024advancing}.
\end{itemize}

The validation activities result in the generation of emissions, which are recorded in the provenance ledger. The carbon budget manager uses the measured values for the purpose of assessing the sustainability thresholds. The Green Validation Orchestrator dynamically adjusts validation depth, model selection, and execution timing based on remaining budget and governance risk thresholds. 


\subsection{Governance Policies}

Building on the proposed architectural extensions, CAGG operationalizes sustainability governance through a set of enforceable policy mechanisms. These policy mechanisms govern validation intensity, regeneration, scheduling decisions, and model selection to ensure that governance assurance objectives are balanced with carbon efficiency constraints.

\begin{itemize}

\item \textbf{Model Escalation Policy:} The validation process begins with lower-compute models, escalating only when the risk of failure remains high. This policy enables proportional allocation of computational resources based on governance risk signals, which is achieved through dynamic model selection by the Green Validation Orchestrator.

\item \textbf{Regeneration Cap Policy:} The number of regeneration iterations is constrained to prevent marginal validation assurance from incurring disproportionate carbon costs. Once regeneration limits are reached, governance gates can trigger human justification or termination of validation iterations. This policy is enforced by the Green Validation Orchestrator, with human approval invoked through governance oversight once the regeneration limit is reached.

\item \textbf{Carbon-Intensity Scheduling Policy:} Schedule computationally intensive validations when the carbon footprint of the grid is relatively smaller, a strategy shown to be feasible in the context of carbon-aware computing\cite{radovanovic2022carbon}. Carbon intensity signals are used to optimize validation scheduling, ensuring environmentally friendly validation processes.

\item \textbf{Budget-Bound Validation Policy:} The scope and depth of validation processes are constrained by carbon budgets and associated governance thresholds. As carbon budgets are depleted, validation processes can be escalated for approval, ensuring that they are within carbon constraints. This policy is enforced by the Carbon Budget Manager in collaboration with governance validation processes.
\end{itemize}

Collectively, these policies are enforced through the Green Validation Orchestrator and Carbon Budget Manager, while the Energy and Carbon Provenance Ledger provides traceable sustainability data, enabling coordinated and auditable governance decisions.

\section{Architectural Design Patterns}

To operationalize CAG within AI-assisted software development, we propose four reusable architectural design patterns. These design patterns have extended the traditional governance mechanisms into the sustainability dimension by leveraging the proposed architectural extensions.

\subsection{Budgeted Governance Gate:} 
This pattern enforces carbon budgets at governance checkpoints to bound the cost of validation and regeneration.
\begin{itemize}
    \item \textbf{Context:} Applied at approval checkpoints such as validation of pull requests, Continuous Integration pipeline stages, or release reviews.
    \item \textbf{Problem:} Validation may involve unbounded computation and carbon expense due to repeated regeneration and/or extensive testing cycles.
    \item \textbf{Solution:} The governance action consumes carbon budget tokens via the proposed Carbon Budget Manager. If the carbon budget is depleted, the gate may deny the request, request a re-attempt in a lower-carbon window, or request human review. This ensures that sustainability constraints are enforced at operational decision points.
\end{itemize}

\subsection{Two-Phase Carbon-Aware Validation}  This pattern runs lightweight checks first and triggers expensive validation only when thresholds are exceeded.
\begin{itemize}
    \item \textbf{Context:} Applied in validation pipelines where multiple levels of testing or verification are available.
    \item \textbf{Problem:} Excessive validation may run unnecessarily, resulting in wasted carbon resources when lightweight validation would have been sufficient.
    \item \textbf{Solution:} The proposed Green Validation Orchestrator performs lightweight validation first, which is followed by a computationally expensive analysis step. Only if certain predefined thresholds are exceeded is a deeper validation phase triggered. This approach reduces unnecessary emissions while maintaining the quality of the validation process.
\end{itemize}
\subsection{Carbon Provenance Evidence} 
This pattern attaches energy and carbon metadata to governance logs to support auditability and reporting.
\begin{itemize}
    \item \textbf{Context:} Applied in governance audit, compliance reports, and post-hoc accountability processes.
    \item \textbf{Problem:} Sustainability impacts of governance activities are often untraceable, which limits transparency and verification of sustainability policies.
    \item \textbf{Solution:} The proposed Energy and Carbon Provenance Ledger provides auditable evidence that allows for the analysis of carbon-aware decisions.
\end{itemize}

\subsection{Stop-and-Justify Regeneration Loop} 
This pattern caps regeneration loops and requires human justification to continue beyond a set limit. Approval from a human in the loop for the repeated regeneration of the outputs of the generative AI.
\begin{itemize}
    \item \textbf{Context:}  Applied in iterative generative AI workflows that require repeated regeneration.
    \item \textbf{Problem:} Excessive regeneration cycles can significantly increase computational and carbon costs without proportional assurance gains.
    \item \textbf{Solution:} After a fixed number of regeneration iterations, the gate pauses the workflow and requires human justification before additional regeneration is allowed.
\end{itemize}

\section{Discussion}



The proposed CAGG shifts sustainability from an infrastructure-centric optimization problem to an architectural concern embedded within governance control mechanisms of GenAI-enabled software development. Existing sustainable AI efforts primarily target model efficiency, hardware optimization, and data center energy management. In contrast, CAGG identifies governance checkpoints such as pull request validation, CI/CD stages, regeneration policies, and release approvals as architectural leverage points where carbon-aware constraints can be enforced alongside assurance and compliance requirements. By embedding carbon budgets, validation orchestration policies, and provenance logging into these control points, sustainability becomes an explicit design parameter influencing validation depth, regeneration frequency, model selection, and execution timing.

From a quality attribute perspective, introducing carbon-aware governance exposes explicit trade-offs that must be architecturally managed. Increasing validation depth or regeneration cycles may improve assurance and compliance confidence, but it increases energy consumption and carbon footprint. Conversely, strict carbon budgets may limit validation rigor in certain contexts. Scheduling compute-intensive validation during lower-carbon time windows may reduce emissions but increase latency and delay developer feedback. Provenance logging improves transparency and auditability, yet adds telemetry and storage overhead. These tensions illustrate that sustainability must be treated as a co-equal architectural quality attribute alongside assurance, performance, and reliability rather than as a secondary optimization objective.

Operationally, CAGG can be integrated into existing DevOps ecosystems with limited structural disruption. Carbon-aware policies can be embedded within existing CI/CD gates, where validation steps are already orchestrated and enforced. For example, a pull request that triggers AI-based test generation and compliance checks may be assigned a predefined carbon budget. The validation orchestrator can execute lightweight checks first and escalate to deeper, model-intensive validation only when risk thresholds justify additional expenditure. If the allocated carbon budget is exceeded, the governance gate may defer execution, downgrade to a lower-energy model, or escalate to human review. The Energy and Carbon Provenance Ledger records inference events and estimated emissions as part of audit logs, enabling traceable sustainability reporting without altering the fundamental structure of the development pipeline.

To reason about the balance between assurance and environmental cost, we introduce an Assurance-per-Carbon perspective, which conceptually reflects the relationship between validation confidence gained and carbon expenditure incurred. While empirical calibration of such a metric requires future study, the perspective provides a structured way to reason about multi-objective trade-offs in governance orchestration decisions. Rather than maximizing validation depth unconditionally, governance policies can aim to optimize assurance gains relative to carbon cost within acceptable risk boundaries.

The approach is subject to several limitations. Carbon footprint estimation depends on available telemetry, hardware characteristics, and regional carbon intensity data, which introduce approximation uncertainty. Overly restrictive budgets may not be appropriate in safety-critical or highly regulated systems where assurance requirements dominate sustainability considerations. In addition, introducing carbon-aware orchestration and provenance logging increases architectural complexity and requires careful policy calibration to avoid excessive workflow friction. Despite these constraints, CAGG demonstrates that governance layers represent actionable and technically feasible intervention points for embedding sustainability into GenAI-enabled software development architectures, thereby extending green software architecture principles to the decision-making layer of AI-assisted engineering workflows.

Sustainability in governance introduces a shift in green software architecture that focuses on sociotechnical control design instead of infrastructure optimization. While prior green AI initiatives focus on the efficiency of AI models and data centers, the proposed approach emphasis on governance layers as leverage points for reducing the environmental impact of GenAI-enabled development ecosystems.

From an architectural perspective, the incorporation of carbon awareness in the governance gates enables sustainability to influence the validation depth, regeneration rates, and model selection in AI models. This shifts sustainability from an operational afterthought to an explicit design concern of the SDLC control mechanisms.

In the implementation of the CAG, there is a need to integrate it in the CI/CD pipelines, ensure it is consistent with the organizational policies, and have access to the data regarding the carbon footprint. In the process, there is a need to ensure a balance between the sustainability and assurance considerations, especially in a high-risk development process where the validation levels cannot be easily reduced.

However, there are also some limitations in the proposed approach, as it could compromise the assurance levels if the carbon footprint is too high, and it could compromise the sustainability levels if the validation levels are too deep. The Assurance-per-Carbon perspective provides a structured way to reason about this tension, though empirical validation remains a subject of future work.

\section{Conclusion}

Governance layers are critical for trustworthy GenAI adoption, but they also pose latent sustainability costs associated with validation, regeneration, and governance activities. As AI-enabled software development ecosystems expand, the sustainability consequences of such workflows become increasingly significant.
This paper proposes Carbon Aware Governance Gates (CAGG), a layered software architecture that integrates sustainability into AI-based governance for GenAI-assisted software development. We present three architectural extensions: Energy and Carbon Provenance Ledger, Carbon Budget Manager, and the Green Validation Orchestrator. These components are accompanied by corresponding governance policies and reusable design patterns. Together, these mechanisms enable governance assurance to be balanced with carbon accountability across SDLC workflows.


\textbf{Acknowledgments}.  This work has been supported by FAST, the Finnish Software Engineering Doctoral Research Network, funded by the Ministry of Education and Culture, Finland, and Business Finland (project ANSE, 1822/31/2025).

\bibliographystyle{IEEEtran}
\bibliography{references}

\end{document}